\documentclass[12pt]{article}
\usepackage[usenames]{color}
\usepackage{graphicx, subfigure}
\usepackage{amsmath, amsthm, amssymb}
\usepackage{amsfonts}
\usepackage{fullpage}
\usepackage{ifthen}
\usepackage{url}
\usepackage[sort&compress]{natbib}
\usepackage{multirow}
\usepackage[linesnumbered,algoruled,boxed,lined,commentsnumbered]{algorithm2e}
\usepackage{bm}
\usepackage{tikz-qtree}
\usepackage{tikz}
\usetikzlibrary{automata,arrows,positioning,calc}
\usepackage{hvlogos}

\makeatletter
\def\@biblabel#1{}
\makeatother

\theoremstyle{plain}
\newtheorem{thm}{Theorem}

\theoremstyle{definition}

\theoremstyle{remark}

\newcommand{\E}{\mathsf{E}}

\newcommand{\prob}{\mathsf{P}}

\newcommand{\pl}{\mathsf{pl}}

\newcommand{\norm}{{\sf N}}

\newcommand{\gam}{{\sf Gamma}}

\newcommand{\Pareto}{{\sf Pareto}}

\newcommand{\Bern}{{\sf Bern}}

\title{A new strategy for finite-sample valid prediction of future insurance claims in the regression setting}

\author{Liang Hong\footnote{Department of Mathematical Sciences, The University of Texas at Dallas, 800 West Campbell Road, Richardson, TX 75080, USA. Tel.:~+1 (972) 883-2161. Email address: liang.hong@utdallas.edu.}}
\date{\today}

\begin{document}

\maketitle

\begin{abstract}
The extant insurance literature demonstrates a paucity of finite-sample valid prediction intervals of future insurance claims in the regression setting.  To address this challenge, this article proposes a new strategy that converts a predictive method in the unsupervised iid (independent identically distributed) setting to a predictive method in the regression setting.   In particular, it enables an actuary to obtain infinitely many finite-sample valid prediction intervals in the regression setting.  

\smallskip

\end{abstract}

\section{Introduction}

The task of predicting future insurance claims is closely related to several key aspects of an insurer's business, such as premium calculation, reserves estimation,  and regulatory compliance.  Therefore,  it is one of the most important tasks actuaries face in their daily work.  When available data consist solely of past claim amounts, this task can only be done in the unsupervised iid  setting.  However,  if data contain information about the claim amount and some predictors/explanatory variables,  actuaries should utilize all available information and treat the task in the regression setting.  In practice,  either case can occur.  Hence, actuaries should be prepared for both.

Broadly speaking, there are two main goals in data science and statistics: (1)~to explain and (2)~to predict (Tukey 1962, p.13; Shmueli 2010).  These two goals are different.  In the regression setting, if the main goal is to explain how explanatory variables affect the response variable, then a model must be close to the true data-generating mechanism.  However, if the chief purpose is prediction,  a model does not need to be close to the true data-generating mechanism: a wrong model can outperform the true data-generating mechanism in some cases (e.g., Shmueli 2010).  In fact,  a model is not even needed to perform prediction (e.g., Hong 2026). 

In the past two decades a plethora of parametric predictive models have been proposed,  see, for instance, Brazauskas and Kleefeld (2011, 2016),  Calder\'{\i}n-Ojeda and Kwok (2016),  Cooray and Ananda (2005),  Frees et al. (2014), Nadarajah and Bakar (2014),  Pigeon and Denuit (2011), and Scollnik (2007).  Two potential issues of a parametric model is (i)~ model misspecification (Hong and Martin 2020) and (ii) selection effect (Hong et al. 2018a, b).  These two concerns prompted researchers to seek some non-parametric predictive models; see, for example,  Fellingham et al. (2015), Hong and Martin (2017),  Jeon and Kim (2013),   Lee and Lin (2010), and Richardson and Hartman (2018).  While a non-parametric predictive model generally does not suffer from the above two issues, they usually have some tuning parameters, whose choices are often subject to debate.  Note that if a data-driven method, such as cross-validation, is applied to choose a tuning parameter, then the effect of selection will result.  Moreover,  prediction based on a non-parametric model is only asymptotically valid, not finite-sample valid; see Section~2.3 for a precise definition of finite-sample validity.  In practice, finite-sample validity is more desirable than asymptotic validity, since no practical algorithm can run forever.  

Methods for finite-sample valid prediction have been developed in the insurance literature (e.g., Hong and Martin 2021; Hong 2026). Though these two methods are both based on conformal prediction---a general machine learning strategy,  they differ in one important aspect: the method in Hong and Martin (2021) is designed for the unsupervised iid setting, whereas the method in Hong (2026) is developed for the regression setting.  Of course,  an actuary can disregard the information supplied by the explanatory variables and apply the method in Hong and Martin (2021) to the data on the response variable to perform prediction in the regression setting.  But doing so might not be optimal, since valuable information provided by the explanatory variables is discarded without reason; see Examples~1--3 in Section~4 for some concrete examples.  Besides the method proposed by Hong (2026),  no other method in the current insurance literature allows actuaries to perform finite-sample valid prediction in the regression setting.  This paucity of methods for finite-sample valid prediction in the regression setting inspires the research presented in this article.  The key purpose of this article is to introduce a new strategy that enables actuaries to apply a predictive method for the unsupervised iid setting to perform prediction in the regression setting, without losing the information supplied by the explanatory variables.  In particular, it leads to infinitely many finite-sample valid prediction intervals for future insurance claims.

The remainder of the paper proceeds as follows.  Section~2 provides the background.  Besides reviewing several key concepts and establishing notational conventions, it defines the problem of predicting future insurance claims in the regression setting and discusses some major approaches to this problem.  Section~3 details the proposed strategy.  Section~4 gives several numerical examples to demonstrate the excellent performance of the new strategy.  Section~5 concludes the article with some remarks. The Appendix reviews conformal prediction and elaborates on a subtle issue in the extant literature.

\section{Preliminaries}

\subsection{The problem at hand}

Suppose the \emph{true data-generating mechanism} is
\begin{equation}
\label{eq:true}
Y=f^\star(X)+\varepsilon, 
\end{equation}
where $Y$ is the response variable,  $f^\star$, often called the \emph{regression function},  is an unknown real-valued function, $X$ is a vector of $p$ predictors, $p\geq 1$, and $\varepsilon$ is a random error term with $\E[\varepsilon]=0$.  Since $Y$ denotes the insurance claim amount,  throughout we assume $Y\geq 0$ unless otherwise specified.  Let $Z_1 = (X_1, Y_1), Z_2=(X_2, Y_2), \ldots, Z_n=(X_n, Y_n)$ be a sequence of iid observations from the data-generating mechanism in (\ref{eq:true}), where $n$ is the sample size.  We are interested in predicting the next response $Y_{n+1}$ at a randomly sampled feature $X_{n+1}$, based on past observations of $Z^n = \{Z_1,\ldots,Z_n\}$.  For this purpose, we can either perform point prediction or interval prediction.  Compared to point prediction,  interval prediction has two advantages. First, it can quantify prediction accuracy in terms of probabilities. Second, it allows the possibility of finite-sample validity.  In this article, we only consider interval prediction.  Specifically, we are interested in the problem of creating a $100(1-\alpha)\%$ prediction interval of  $Y_{n+1}$ at a randomly sampled feature $X_{n+1}$, based on past observations of $Z^n$, where $0<\alpha<1$. Since $Y\geq 0$,  we stipulate that this prediction interval must be of the form $[0, u(X_{n+1}, Z^n))$, where $u$ is a functional of $(X_{n+1}, Z^n)$.  Note that here the upper bound $u(X_{n+1}, Z^n)$ is random.  Hence, the prediction interval $[0, u(X_{n+1}, Z^n))$ a random prediction interval. One can also use a deterministic prediction interval.  For example,  if $u_{\alpha}$ is the $100(1-\alpha)\%$-th quantile of $Y$, then $[0,  u_{\alpha})$  is a deterministic prediction interval for $Y_{n+1}$. Indeed, it is the shortest deterministic prediction interval of the form $[0, a)$ with a coverage probability of $1-\alpha$.  We will refer to  $[0,  u_{\alpha})$ as the \emph{oracle $100(1-\alpha)\%$ prediction interval} and its length the \emph{oracle length}.  It usually serves as the benchmark for assessing the efficiency of other prediction intervals.  One key advantage of a random interval is that it can be more efficient than the oracle prediction interval in some cases; we will demonstrate this in Section~4.

\subsection{Parametric models, non-parametric models, and model-free \\ methods}

Generally speaking, there are two broad approaches to the aforementioned problem:  the model-based approach and the model-free approach.  In the model-based approach, we first posit a \emph{(predictive) model}
\begin{equation*}
\mathcal{M}=\{(\mathcal{F},  \mathcal{D})\},
\end{equation*}
where $\mathcal{F}$ is a class of real-valued functions and $\mathcal{D}$ is a family of distributions.  That is, a model accounts for two sources of uncertainty in (\ref{eq:true}): the form of the regression $f^\star$ and the distribution of $\varepsilon$.  For example, in the classical linear model, we have
\begin{eqnarray*}
\mathcal{F} &=& \{f(t_1, \ldots, t_p)=a_0+a_1t_1+\ldots+a_pt_p\mid   \text{ $a_i\in\mathbb{R}$ for $0\leq i\leq n$} \}, \\
\mathcal{D} &=& \{\norm(0, \sigma^2)\mid \sigma>0\},
\end{eqnarray*}
where $\norm(\mu, \sigma^2)$ stands for the normal distribution with mean $\mu$ and variance $\sigma$.
If both $\mathcal{F}$ and $\mathcal{D}$ can be characterized by finitely many parameters,  $M$ is called a \emph{parametric model}; otherwise, we say $M$ is a \emph{non-parametric model}.  If $f^\star\in\mathcal{F}$ and the distribution of $\varepsilon$ belongs to $\mathcal{D}$,  we say the model $\mathcal{M}$ is \emph{well-specified} or \emph{correct}; otherwise, we say the model $\mathcal{M}$ is \emph{misspecified} or \emph{wrong}.  For example,  when $f^\star(t_1, \ldots, t_p)=t_1+\ldots+t_p$, the classical linear model is correct if $\varepsilon\sim \norm(0, 1)$, but it will be wrong if $\varepsilon$ follows the $t$-distribution.  Since neither $f^\star$ nor the distribution of $\varepsilon$ is observable, there is no way to be certain that a model $\mathcal{M}$ is correct even if it is.  Though a wrong model can sometimes outstrip the true data-generating mechanism (Shmueli 2010),  it often leads to misleading predictions (e.g., Hong 2026). Therefore,  actuaries should take model misspecification risk into account when they employ the model-based approach.  There is another issue, called the \emph{selection effect},  that is associated with the model-based approach. This issue occurs when a model selection tool is first used to select a model, and then predictions are performed using the chosen model.  Each step of this two-step procedure,  examined alone, is beyond reproach. However,  when we combine them in practice,  serious biases can ensue.  The reason is that the model is treated as random (because it is data-dependent) during this process; however, the textbook formula for prediction presumes the model is fixed (not data-dependent).  For a parametric model,  a robust device might reduce the model misspecification risk, but no satisfactory method is known at this point to treat the selection effect (e.g., Kuchibhotla et al. 2022).  Generally speaking, non-parametric models are not susceptible to the model misspecification risk because these models all approximate the regression function $f^\star$ well, irrespective of the form of $f^\star$.  As for the selection effect, the matter is more subtle.  Nearly all existing non-parametric models have tuning parameters. If these parameters are chosen with a data-driven method, such as cross-validation, then the selection effect will be present. If the tuning parameters are chosen subjectively,  the selection effect is avoided, but the choice of each parameter can be questionable. 

It is evident from the above discussion that both model misspecification and selection effect stem from the fact that we have to pin down a model in the model-based approach.  Therefore,  a natural way to circumvent the difficulties posed by model misspecification and selection effect is to avoid using any model.  This is the philosophy of the model-free approach.  The model-free approach applies to both estimation and prediction. It is not foreign to actuaries.  For example,  using sample means to estimate the population mean is a model-free method, though it falls within the ambit of estimation.  As for prediction,  decision trees, $K$-nearest neighbors (KNN),  and random forests are familiar examples of model-free methods.  Figure~1 illustrates the key difference between a model-based method and a model-free method for prediction.  In a model-based method,  model training is a necessary step before prediction, while in a model-free method,  no model ever enters the scene.


\begin{center}
	\begin{tikzpicture}[->, >=stealth', auto, semithick, node distance=3.5cm]
	\tikzstyle{roundnode}=[fill=white,draw=black,thick,text=black,scale=2.0]	
	\node[roundnode]    (A)                     		{Data};     
	\node[roundnode]    (B) [above right of=A]   {Predictive Models};
	\node[roundnode]    (C)[below right of=B] {Prediction};
	\path
	(A) edge[left,above]			node{model-free prediction}      (C)
	      edge[left,above]	                  node[rotate=45]{model building}	(B)
	(B) edge[right, above]                 node[rotate=-45]{model-based prediction}       (C);
	\end{tikzpicture}          
\end{center}
\begin{center}
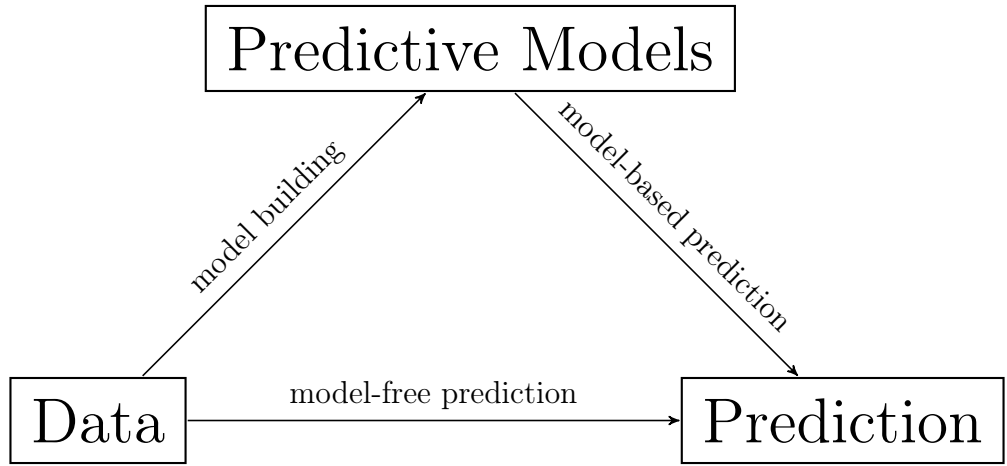
	Figure~1: Model-based methods versus model-free methods for prediction. 
\end{center}

\bigskip

Note that both a non-parametric model and a model-free method are under the umbrella of non-parametric statistics, though the former is model-based and the latter is model-free.   Figure~2 illustrates the classification of different predictive methods.

\bigskip

\begin{center}
	\begin{tikzpicture}
[
    level 1/.style={sibling distance=55mm},
    level 2/.style={sibling distance=55mm},
]
	\node [font=\large]{Predictive methods}
		child {node [font=\large]{Parametric models}}
		child {
		    node [font=\large] {Non-parametric methods}
		    child {node [font=\large]{Non-parametric models}}
		    child {node [font=\large]{Model-free methods}}
		};
\end{tikzpicture}
\begin{center}
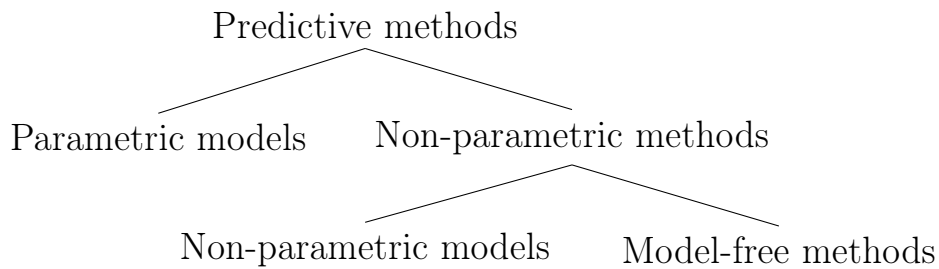
\title{Figure~2: Classification of predictive methods.}
\end{center}
\end{center}

\bigskip

Figure~3 provides an alternative classification of predictive methods.  Note that a non-parametric model falls within the purview of model-based methods. 

\bigskip
	
\begin{center}
	\begin{tikzpicture}
[
    level 1/.style={sibling distance=55mm},
    level 2/.style={sibling distance=55mm},
]
	\node [font=\large]{Predictive methods}
		child {
			    node [font=\large] {Model-based methods}
			    child {node [font=\large]{Parametric models}}
			    child {node [font=\large]{Non-parametric models}}
			}
		child {node [font=\large]{Model-free methods}}
		;
\end{tikzpicture}
\begin{center}
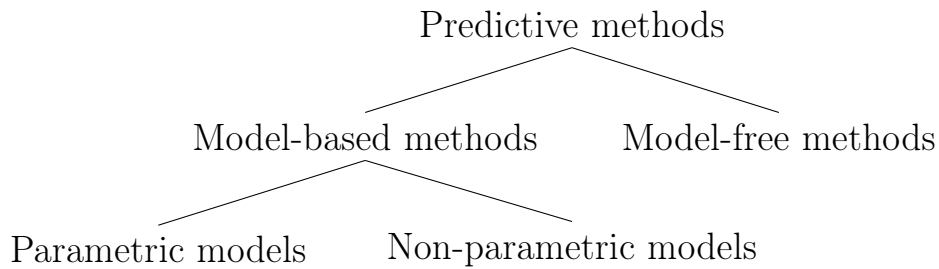
\title{Figure~3: Alternative classification of predictive methods.}
\end{center}
\end{center}

\subsection{Two types of validity}

Suppose $\prob_Z$ is the distribution of $Z_1=(X_1, Y_1)$. For $0<\alpha<1$, a $100(1-\alpha)\%$ prediction interval $I_{\alpha}(X_{n+1}, Z^n)$ is said to be \emph{asymptotically valid} if 
\begin{equation*}
\lim_{n\rightarrow \infty} \prob^{n+1}_Z \{Y_{n+1}\in I_{\alpha}(X_{n+1}, Z^n)\}\geq 1-\alpha, \quad \text{for all $\prob_Z$},
\end{equation*}
where $n$  is the sample size and $\prob^{n+1}_Z$ denotes the joint distribution for $Z_1$, $\ldots, Z_n$, $Z_{n+1}$.  Intuitively,  the coverage probability of an asymptotically valid $100(1-\alpha)\%$ prediction interval will reach the confidence level $1-\alpha$ when the sample size $n$ goes to infinity.  Prediction intervals based on most non-parametric methods are asymptotically valid.  However,  all samples are finite in actuarial practice.  Hence, we want the coverage probability of the prediction interval $I_{\alpha}(X_{n+1}, Z^n)$ to be at least $1-\alpha$ for any finite sample size $n$.  We say a $100(1-\alpha)\%$ prediction interval $I_{\alpha}(X_{n+1}, Z^n)$ is \emph{finite-sample valid} if 
\begin{equation}
\label{eq:valid}
\prob^{n+1}_Z\{Y_{n+1}\in I_{\alpha}(X_{n+1}, Z^n)\}\geq 1-\alpha, \quad \text{for all $n$ and all $\prob_Z$}.
\end{equation}
Note that a finite-sample valid prediction interval is automatically asymptotically valid.  In addition, finite-sample validity guarantees that the prediction interval $I_{\alpha}(X_{n+1}, Z^n)$ achieves the advertised confidence level for all sample sizes $n$, regardless of the distribution $\prob$.  Therefore,   a finite-sample prediction interval cannot be created using a parametric method.  Though finite-sample validity seems to be a lofty goal,  it can be achieved using a general machine learning strategy called \emph{conformal prediction}.  For a general treatment of conformal prediction, see Vovk et al.  (2005) and Shafer and Vovk (2008);  for applications of conformal prediction in insurance,  see Hong and Martin (2021) and Hong (2026).  In the regression setting,  many conformal prediction intervals have been proposed (e.g., Lei et al. 2013, Lei and Wasserman 2014; Lei et al.  2018).  Though these prediction intervals are finite-sample valid in theory,  none of them can be determined exactly in practice.  Authors of these prediction intervals often come up with an approximation.  This seemingly innocent practice has a devastating effect: the finite-sample validity of the approximated prediction interval is nowhere justified; see the Appendix for a detailed discussion of this serious issue.  Therefore,  in the regression setting,  barely any usable finite-sample valid prediction interval of the form $[0, a)$ is available to practicing actuaries, except the one in Hong (2026).  Finally,   it is worth remarking that a model-free predictive method avoids the issues of model misspecification and selection effect, but it generally does not guarantee finite-sample validity.  At the time of writing, conformal prediction is the only known (model-free) method for finite-sample valid prediction. 

Table~\ref{table:models} summarizes three major approaches to constructing prediction intervals with respect to two potential issues (model misspecification and selection effect) and two desirable properties (asymptotic validity and finite-sample validity).

\begin{table}[!ht]
\begin{center}
\begin{tabular}{l|ccc}
\hline
    			   & Parametric models & Non-parametric models  & Conformal prediction \\
\hline
Model misspecification         &  Yes   				    & Yes                      & No   \\
Selection effect              	   &  Yes   				    & Maybe 		& No  \\
Asymptotic validity    	   &  No  				    & Yes			& Yes \\
Finite-sample validity    	   &  No  				    & No  			& Yes \\
\hline
\end{tabular}
\end{center}
\caption{Comparison of three major approaches to constructing prediction intervals with respect to model misspecification, selection effect,  asymptotic validity, and finite-sample validity}
\label{table:models}
\end{table}

\section{New strategy}

\subsection{Proposed strategy}

To create more finite-sample valid prediction intervals in the regression setting,  we first consider a new strategy that converts a prediction interval in the unsupervised iid setting to a prediction interval in the regression setting.  This strategy is applicable in a general context.  Thus,  here we first describe the proposed strategy without the constraint $Y\geq 0$.  In the next section, we will show how to tailor it to the problem of predicting future insurance claims. 

First,  we take a $p$-variate real-valued  function $h$ and write (\ref{eq:true}) as 
\begin{eqnarray}
\label{eq:true2}
Y &=& h(X)+[f^\star(X)+\varepsilon]-h(X) \nonumber\\
  &=&   h(X)+[Y-h(X)] \nonumber \\
   &=&  h(X)+W,
\end{eqnarray}
where $W=Y-h(X)$.  We will refer to the function $h$ as a \emph{transformation}.  Its choice is at the discretion of the actuary.  Put $W_i=Y_i-h(X_i)$ for $i\geq 1$ and $W^n=\{W_1, \ldots, W_n\}$. Since $(X_1, Y_1), \ldots, (X_n, Y_n), \ldots$ is a sequence of iid random vectors, $W_1, \ldots, W_n , \ldots$ is a sequence of iid random variables.  Next, we construct a $100(1-\alpha)\%$ prediction interval $(L(W^n), U(W^n))$ for $W_{n+1}$, where $L(W^n)$ and $U(W^n)$ are the lower bound and upper bound respectively.  (Note that this step requires a method for constructing a prediction interval in the unsupervised iid setting.) Since $W_{n+1}=Y_{n+1}-h(X_{n+1})$,  we have
\[
L(W^n)<W_{n+1}<U(W^n) \text{ if and only if $L(W^n)+h(X_{n+1})<Y_{n+1}<U(W^n)+h(X_{n+1})$}.
\]
Finally,  we obtain a $100(1-\alpha)\%$ prediction interval for $Y_{n+1}$ as 
\begin{equation}
\label{eq:PI_general}
(L(W^n)+h(X_{n+1}), U(W^n)+h(X_{n+1})).
\end{equation}

If in addition the $100(1-\alpha)\%$ prediction interval $(L(W^n), R(W^n))$ is finite-sample valid,  i.e., 
\begin{equation*}
\prob^{n+1}_W\{L(W^n)<W_{n+1}< U(W^n) \}\geq 1-\alpha\quad \text{for all $n,$ and all $\prob_W$},
\end{equation*}
where $\prob_W$ denotes the distribution of $W_1$ and $\prob^{n+1}_W$ stands for the joint distribution for $W_1, \ldots, W_n, W_{n+1}$, then
\begin{equation*}
\prob^{n+1}_Z\{L(W^n)+h(X_{n+1})< Y_{n+1}<U(W^n)+h(X_{n+1}\} \}\geq 1-\alpha, \quad \text{for all $n$ and all $\prob_Z$}.
\end{equation*} 
In this case,  the $100(1-\alpha)\%$ prediction interval given by (\ref{eq:PI_general}) is finite-sample valid for any $h$.

There are numerous available methods for constructing a $100(1-\alpha)\%$ prediction interval in the unsupervised iid setting (e.g., Tian et al 2022).  Each of them can be used to construct $(L(W^n), U(W^n))$.  However,  the only one known to be finite-sample valid (e.g.,  Frey 2013; Hong and Nasreddine 2025) is
\[
(W_{(l)}, W_{(r)}), 
\]
where $1\leq l<r\leq n$, $(r-l)/(n+1)\geq 1-\alpha$ (e.g., $l=\min\{n, \lfloor (n+1)(\alpha/2)\rfloor+1\}$ and $r=(n+1)-l$). 
With this choice for  $(L(W^n), U(W^n))$,  our proposed prediction interval in  (\ref{eq:PI_general}) becomes
\begin{equation*}
(W_{(l)}+h(X_{n+1}), W_{(r)}+h(X_{n+1})).
\end{equation*}

\subsection{Application to prediction of insurance claims}

To apply the proposed strategy to predict future insurance claims,  we must overcome additional challenges caused by the constraint $Y\geq 0$.  We still first choose a ($p$-variate real-valued) transformation $h$  and write $Y=h(X)+Y$ as in (\ref{eq:true2}), except that we require $h(t)\geq 0$ for all $t\in\mathbb{R}^p$.  Next, we construct a one-sided $100(1-\alpha)\%$ prediction interval $(-\infty,  b(W^n))$ of $W_{n+1}$ at confidence level $1-\alpha$, where $b(W^n)$ is the upper bound of this one-sided prediction interval for $W_{n+1}$.  (Note that $W_{n+1}$ is not necessarily non-negative.) Since $0\leq Y_{n+1}$ and $W_{n+1}=Y_{n+1}-h(X_{n+1})$,
\[
W_{n+1}\leq b(W^n) \text{ if and only if $0\leq Y_{n+1}\leq b(W^n)+h(X_{n+1})$}.
\]
The last equation suggests that $[0, b(W^n)+h(X_{n+1}))$ might serve as a $100(1-\alpha)\%$ prediction interval for $Y_{n+1}$.  However,  there is a nuisance: $b(W^n)+h(X_{n+1})$ might not be positive for a given random sample of $(X, Y)$.  Since $h(X_{n+1})$ is non-negative, $b(W^n)+h(X_{n+1})<0$ implies $b(W^n)<0$. Also, $W_{n+1}<0$ implies $Y_{n+1}<h(X_{n+1})$. Thus, we can construct a $100(1-\alpha)\%$ prediction interval for $Y_{n+1}$ as follows:
\begin{equation}
\label{eq:PI_special1}
\left\{
                           \begin{array}{ll}
                           \lbrack 0, b(W^n)+h(X_{n+1})),& \hbox{if $b(W^n)+h(X_{n+1})>0$,}  \\
                           \lbrack 0, h(X_{n+1})), & \hbox{if $b(W^n)+h(X_{n+1})\leq 0$.} 
                          \end{array}
                         \right.
\end{equation}
This prediction interval has two drawbacks. First, $h(t)$ can be zero for some $t$, rendering the prediction interval degenerate.  Secondly, if $h(t)$ is large for all $t$, then the prediction interval may be ``too long''.  Therefore, instead of using the prediction interval in (\ref{eq:PI_special1}), we shall consider the following $100(1-\alpha)\%$ prediction interval for $Y_{n+1}$:
\begin{equation}
\label{eq:PI_special2}
\left\{
                           \begin{array}{ll}
                           \lbrack 0, b(W^n)+h(X_{n+1})),& \hbox{if $b(W^n)+h(X_{n+1})>0$,}  \\
                           \lbrack 0, \min\{u(Y^n),  h(X_{n+1})\}), & \hbox{if $b(W^n)+h(X_{n+1})\leq 0$,} 
                          \end{array}
                         \right.
\end{equation}
where $u(Y^n)$ is the upper bound of  a $100(1-\alpha)\%$ prediction interval $[0, u(Y^n))$ for $Y$, based only on $Y^n$.

A favored choice of $(-\infty, b(W^n))$ is the model-free $100(1-\alpha)\%$ prediction interval $(-\infty, W_{(r)}))$, where $r=\min\{n, \lfloor(n+1)(1-\alpha \rfloor+1\}$), and $W_{(k)}$ is the $k$-th order statistics of $W_1, \ldots, W_n$.  This prediction interval is finite-sample valid (e.g.,  Frey 2013; Hong and Nareddine 2025).  That is, 
\begin{equation*}
\prob^{n+1}_W\{W_{n+1}\leq W_{(r)} \}\geq 1-\alpha\quad \text{for all $n,$ and all $\prob_W$}.
\end{equation*}
Similarly, a preferred choice of $[0, u(Y^n))$ is $[0, Y_{(r)})$.  This prediction interval is also finite-sample valid (e.g., Hong and Martin 2021),  namely, 
\begin{equation*}
\prob^{n+1}_Y\{Y_{n+1}\leq Y_{(r)} \}\geq 1-\alpha\quad \text{for all $n,$ and all $\prob_Y$},
\end{equation*}
where $\prob_Y$ denotes the distribution of $Y_1$ and $\prob^{n+1}_Y$ stands for the joint distribution for $Y_1, \ldots, Y_n, Y_{n+1}$.  
With these choices,  the $100(1-\alpha)\%$ prediction interval for $Y_{n+1}$ given by (\ref{eq:PI_special2}) specializes to
\begin{equation}
\label{eq:intervalpred2}
I_{\alpha}^h(X_{n+1}, Y^n)=
\left\{
                           \begin{array}{ll}
                           \lbrack 0, W_{(r)}+h(X_{n+1})),& \hbox{if $W_{(r)}+h(X_{n+1})>0$;}  \\
                           \lbrack 0, \min\{Y_{(r)},  h(X_{n+1})\}), & \hbox{if $W_{(r)}+h(X_{n+1})\leq 0$.} 
                          \end{array}
                         \right.
\end{equation}
Also,  it is clear that 
\begin{equation*}
\prob^{n+1}_Z\{Y_{n+1}\in I_{\alpha}^h(X_{n+1}, Y^n) \}\geq 1-\alpha, \quad \text{for all $n$ and all $\prob_Z$}.
\end{equation*}
Therefore,  $I_{\alpha}^h(X_{n+1}, Y^n)$  is finite-sample valid for any non-negative transformation $h$.   Since there are infinitely many choices of such an $h$, (\ref{eq:intervalpred2}) immediately yields infinitely many finite-sample valid $100(1-\alpha)\%$ prediction intervals for $Y_{n+1}$.  Note that we do not push further to replace the upper bound $W_{(r)}+h(X_{n+1})$ with $\min \{Y_{(r)}, W_{(r)}+h(X_{n+1})\}$ in (\ref{eq:intervalpred2}).  A close examination of our derivation reveals that such a change will destroy the finite-sample validity of the resulting prediction interval.  

Though $I_{\alpha}^h(X_{n+1}, Y^n)$ is finite-sample valid for any non-negative $h$,  the choice of $h$ deserves some consideration.  First, if $h(t)=0, \ t\in \mathbb{R}^p$, then $I_{\alpha}^h(X_{n+1}, Y^n)$ specializes to $[0, Y_{(r)})$.  Secondly, if $h(t_1, \ldots, t_p)$ tends to increase fast in any of $t_i$ ($i=1, \ldots, p$), then the mean value of $W_{(r)}+h(X_{n+1})$ will likely be large,  leading to a relatively large mean length of $I_{\alpha}^h(X_{n+1}, Y^n)$.  For example, if $p=1$, then the mean length of $I_{\alpha}^h(X_{n+1}, Y^n)$ with $h(t)=t^2$ is expected to exceed that of $I_{\alpha}^g(X_{n+1}, Y^n)$ with $g(t)=t, \ t\geq 0$.  In view of this and the principle of parsimony,  a simple $h$ with a slowly increasing rate in each of its arguments is strongly recommended.   Simulation studies in Section~4 confirm this intuition.  Finally, it is evident from the above derivation that the requirement $h(t)\geq 0$ for all $t\in \mathbb{R}^p$ can be relaxed to the requirement $h(X_1, \ldots, X_p)\geq 0$.

\section{Illustration}

\subsection{Simulated data}

\subsection*{Example 1}

Let $\gam(\alpha, \lambda)$ denote the gamma distribution with shape parameter $\alpha>0$ and rate parameter $\lambda>0$,  i.e., the gamma distribution whose density function is given by 
\[
f(x)=\frac{\lambda^\alpha}{\Gamma(\alpha)}x^{\alpha-1}e^{-\lambda x}, \quad x>0.
\]
Suppose the true data-generating mechanism is 
\[
Y=X+\varepsilon,
\]
where $X\sim \gam(5,4)$,  $\varepsilon\sim \gam(0.5,  4)$,  and $X$ and $\varepsilon$ are independent.  Though $Y$ is linear in $X$,  the fact that $Y\geq 0$ implies that the true data-generating mechanism is not the classical linear model, but a bona fide generalized linear model.  It follows from the basic properties of the gamma distribution that $Y\sim \gam(5.5, 3)$.  Therefore, in this case the oracle $100(1-\alpha)\%$ prediction interval is $[0,  u_{\alpha})$, where $u_{\alpha}$ is the $100(1-\alpha)\%$ quantile of $\gam(5.5, 4)$.  As in previous works on finite-sample valid prediction of future insurance claims (e.g., Hong and Martin 2021; Hong 2026),  we will use the length of this oracle prediction interval (i.e.,  the oracle length) as the benchmark. We consider three different transformations: a linear transformation, a polynomial transformation, and a non-polynomial transformation.  
\begin{eqnarray*}
h_1 (t) &=& t,\\
h_2(t) &=&  t^2+3t, ,\\
h_3(t) &=&  \log(1+t).
\end{eqnarray*}
Then $h_i(X)\geq 0$ for $i=1, 2,3$. For the confidence level $1-\alpha=0.9$,  we generate $N$ random samples of size $n+1$, where $N=3, 000$ and $n=50$.  For each of these samples,  we construct the $100(1-\alpha)\%$ prediction interval $I_{\alpha}^h(X_{n+1}, Y^n)$ in (\ref{eq:intervalpred2}) using the first $50$ observations and the predictor of the last  observation (i.e.,  the $51$st observation), and then test whether the resulting prediction interval contains the last response variable.  We estimate the coverage probability of the $I_{\alpha}^h(X_{n+1}, Y^n)$ as $M/N$, where $M$ is number of times $I_{\alpha}^h(X_{n+1}, Y^n)$ contains the last response variables among these $N$ realizations of $I_{\alpha}^h(X_{n+1}, Y^n)$.  In addition, we calculate the mean length of these $N$ realizations of $I_{\alpha}^h(X_{n+1}, Y^n)$ and compare it to the oracle length.  We also consider the $100(1-\alpha)\%$ prediction interval $[0, Y_{(r)})$ based solely on the $Y$-data and estimate its coverage probability and mean length. Table~\ref{table:ex1} summarizes the results.  

\begin{table}[!ht]
\begin{center}
\scalebox{1.0}{
\begin{tabular}{c|cc} 
\hline
Prediction Interval  	&   Coverage Probability &  Mean Length Relative to the Oracle Length\\
 \hline
 $[0, Y_{(r)})$     		        				& 90\% 		&  1.02\\
$I_{\alpha}^{h_1}(X_{n+1}, Y^n)$     		        & 91\%     		&   0.75 \\
$I_{\alpha}^{h_2}(X_{n+1}, Y^n)$      			& 92\% 	         &   1.36\\
$I_{\alpha}^{h_3}(X_{n+1}, Y^n)$                 	& 91\% 		&   0.89\\
\hline
\end{tabular}
}
\end{center}
\caption{\small {Coverage probabilities and mean lengths relative to oracle length for the $90\%$ prediction intervals in Example~1.}}
\label{table:ex1}
\end{table}

All four prediction intervals achieve their respective nominal coverage level $90\%$, and their performance are comparable in terms of coverage probability.  The mean lengths of $I_{\alpha}^{h_1}(X_{n+1}, Y^n)$ and $I_{\alpha}^{h_3}(X_{n+1}, Y^n)$ are much shorter than the oracle length.  We reiterate that the length of the oracle $100(1-\alpha)\%$  prediction interval is the shortest among all \emph{deterministic} $100(1-\alpha)\%$  prediction intervals.  The mean length of a genuinely random $100(1-\alpha)\%$  prediction interval can be less than it.  Additionally,  the mean lengths of $I_{\alpha}^{h_1}(X_{n+1}, Y^n)$ and $I_{\alpha}^{h_3}(X_{n+1}, Y^n)$ are much shorter than the mean length of $[0, Y_{(r)})$.  This confirms our intuition that information furnished by predictors is generally useful in prediction.  The mean length of $I_{\alpha}^{h_2}(X_{n+1}, Y^n)$ is longer than both the oracle length and the mean length of $[0, Y_{(r)})$.  This is anticipated,  since $h_2$ grows relatively fast in $t$.

\subsection*{Example 2}

Suppose $\gam(\beta, \theta)$ denotes the (type II) Pareto distribution with density function
\[
p(x)=\frac{\beta \theta^\beta}{(\theta+x)^{\beta+1}}, \quad x>0.
\]
We perform the same simulation experiment as in Example~1 with three exceptions. First,  the data is generated from the following mechanism:
\[
Y=X_1+X_2+\varepsilon,
\]
where $X_1\sim \gam(5, 2)$,  $X_2\sim \Pareto(3, 5)$,  $\varepsilon\sim \gam(0.5,  3)$,  and $X_1$, $X_2$,  and $\varepsilon$ are independent.  Second,  we take the following transformations:
\begin{eqnarray*}
g_1 (t_1, t_2) &=& t_1+0.5t_2, \\
g_2(t_1, t_2)&=& (t_1^3+t_2)/2, \\
g_3(t_1, t_2)&=& \log(1+t_1+t_2).
\end{eqnarray*}
Note that $g_i(X_1, X_2)\geq 0$ for $i=1, 2,3$.  Finally, here we do not have a closed-form formula for the upper bound of the oracle $90\%$ prediction interval for $Y$.  Therefore,  we estimate it using the empirical quantile, based on a sample of size $5,000$ that is independent from the $N$ random samples.  Table~\ref{table:ex2} summarizes the results.  

\begin{table}[!ht]
\begin{center}
\scalebox{1.0}{
\begin{tabular}{c|cc} 
\hline
Prediction Interval  	&   Coverage Probability &  Mean Length Relative to the Oracle Length\\
 \hline
 $[0, Y_{(r)})$     		        				& 92\% 		&  1.11\\
$I_{\alpha}^{g_1}(X_{n+1}, Y^n)$     		         & 91\%     		&   0.84 \\
$I_{\alpha}^{g_2}(X_{n+1}, Y^n)$                 	& 91\% 		&   1.87\\
$I_{\alpha}^{g_3}(X_{n+1}, Y^n)$      			& 92\% 	         &   1.03\\
\hline
\end{tabular}
}
\end{center}
\caption{\small {Coverage probabilities and mean lengths relative to oracle length for the $90\%$ prediction intervals in Example~2.}}
\label{table:ex2}
\end{table}

Since the Pareto distribution is heavy-tailed,  the distribution of $Y$ is also heavy-tailed.  Similar to what we observed in Example~1,  all four prediction intervals achieve the nominal coverage level $90\%$.  In terms of coverage probability, they are comparable.  In terms of mean length, $I_{\alpha}^{g_1}(X_{n+1}, Y^n)$ is the best; it is the only one that beats the oracle prediction interval. The mean length of $I_{\alpha}^{g_3}(X_{n+1}, Y^n)$ is almost the same as the oracle length. However,  $I_{\alpha}^{g_2}(X_{n+1}, Y^n)$ is unsatisfactorily conservative with its mean length being $1.87$ times of the oracle length.  Once again, we see that when the transformation does not increase fast in any of its arguments, the proposed method tends to generate an efficient finite-sample valid prediction interval; otherwise, the resulting prediction interval, though still finite-sample valid,  is likely to be conservative. 

\subsection*{Example 3}
Let $\Bern(p)$ denote the Bernoulli distribution with success probability $p$ where $0<p<1$. We run the same simulation as in Example~2 with two exceptions. First, the true data-generating mechanism is
\[
Y=1+X_1+X_2+X_3+\epsilon,
\]
where $X_1\sim \gam(5, 4)$,  $X_2\sim \Pareto(3, 5)$,  $-X_3\sim \Bern(1/3)$, $\varepsilon\sim \gam(0.5,  4)$,  and $X_1$, $X_2$,  $X_3$, and $\varepsilon$ are independent.   Here both categorical and numerical predictors are present. Second, the transforms are taken to be
\begin{eqnarray*}
f_1 (t_1, t_2, t_3) &=& 1+t_1+t_2+t_3,\\
f_2(t_1, t_2, t_3)&=&  (t_1^2+t_2^2+t_3^2)/2,\\
f_3(t_1, t_2,  t_3)&=&  \log(2+t_1+t_2+t_3).
\end{eqnarray*}
Evidently, $h_i(X_1, X_2, X_3)\geq 0$ for $i=1, 2,3$.  The simulation results are summarized in Table~\ref{table:ex3}.  

\begin{table}[!ht]
\begin{center}
\scalebox{1.0}{
\begin{tabular}{c|cc} 
\hline
Prediction Interval  	&   Coverage Probability &  Mean Length Relative to the Oracle Length\\
 \hline
 $[0, Y_{(r)})$     		        				& 90\% 		&  1.06\\
$I_{\alpha}^{f_1}(X_{n+1}, Y^n)$     		         & 90\%     		&   0.62 \\
$I_{\alpha}^{f_2}(X_{n+1}, Y^n)$                 	& 90\% 		&   1.71\\
$I_{\alpha}^{f_3}(X_{n+1}, Y^n)$      			& 90\% 	         &   0.99\\
\hline
\end{tabular}
}
\end{center}
\caption{\small {Coverage probabilities and mean lengths relative to oracle length for the $90\%$ prediction intervals in Example~3.}}
\label{table:ex3}
\end{table}

Similar to what we have observed in Examples~1 and 2,  all four prediction intervals attain the nominal coverage level.  In terms of efficiency,  $[0, Y_{(r)})$ and the proposed prediction interval based on a logarithmic transformation, i.e, $I_{\alpha}^{f_3}(X_{n+1}, Y^n)$,  are both nearly as efficient as the oracle prediction interval.  The proposed prediction interval based on a linear transformation, i.e., $I_{\alpha}^{f_1}(X_{n+1}, Y^n)$,  is much more efficient than the other three prediction intervals.  However,  the proposed prediction interval based on the quadratic polynomial transformation, i.e.,  $I_{\alpha}^{f_2}(X_{n+1}, Y^n)$,  is very conservative.

\subsection{Automobile bodily injury claims data}

Consider a real data set accompanying Frees (2010).  This data set on automobile injury claims is based on a 2002 study from the Insurance Research Council (IRC), a division of the American Institute for Chartered Property Casualty Underwriters and the Insurance Institute of America.  The sample size of the data set is $n = 1, 340$. There are one response variable ``LOSS'' (claim amount in thousands) and seven predictors: ``CASENUM'' (case number), ``ATTORNEY'' (whether the claimant is represented by an attorney), ``CLMSEX'' (claimant's gender),  ``MARITAL'' (claimant's marital status), ``CLMINSUR'' (whether the driver of the claimant's vehicle is insured), ``SEATBELT'' (whether the claimant was wearing a seat belt),  and ``CLMAGE'' (claimant's age).  Missing values are present for several predictors. 
This data set is available at the following website: \url{https://instruction.bus.wisc.edu/jfrees/jfreesbooks/regression%20modeling/bookwebdec2010/data.html} 
Since the case number was assigned after an accident, and most claimants will decide whether or not to retain an attorney after an accident,  we will not use data on these two predictors.  In addition, we replace each missing value with $0$.  We set $Y=$``CLMAGE'',  $X_1=$``CLMSEX'', $X_2=$``MARITAL'',  $X3=$``CLMINSUR'',  $X4=$``SEATBELT'',  and $X5= $``CLMAGE''.   Table~\ref{table:ex4.1} gives the summary statistics of these variables.

\begin{table}[!ht]
\begin{center}
\scalebox{0.9}{
\begin{tabular}{l|ccccccc}
\hline
 Variable       & Minimum & 1st Quartile & Median & 3rd Quartile  & Maximum & Mean\\
 \hline
$Y$               & 0.005     & 0.640           & 2.331     & 3.995       &    1067.697   & 5.954\\
$X1$		     &0.000    & 1.000 	  & 2.000     &  2.000       &   2.000 & 1.545 \\
$X2$	    &0.000    & 1.000 	         & 2.000     &  2.000       &   4.000 & 1.574 \\
$X3$	    &0.000    & 2.000 	         & 2.000     &  2.000       &   2.000 & 1.849 \\
$X4$	   &0.000    & 1.000 	  & 1.000     &  1.000       &   2.000 & 0.981 \\
$X5$	   &0.000    & 16.00 	  & 27.00     &  41.00      &   95.00 & 27.94 \\
\hline
\end{tabular}
}
\end{center}
\caption{\small  Summary statistics of selected variables based on the automobile bodily injury claims data in Example~4.  }
\label{table:ex4.1}
\end{table}
To apply the proposed method, we choose a logarithmic transformation 
\[
h(t_1, t_2, t_3, t_4, t_5)=\log (1+t_1+t_2+t_3+t_4+t_5)
\]
Since all predictors take non-negative values,  $h(X_1, X_2, X_3, X_4, X_5)\geq 0$.  For confidence levels $1-\alpha=0.9$,  $0.925$,  $0. 95$,  $0.975$, we apply our method using the first $1, 339$ data entries and the values of the five predictors of the last (i.e., $1, 340$-th) entry.  We also determine the upper bounds of the oracle prediction interval (based on empirical quantile) and $[0, Y_{r})$ at chosen confidence levels.  Table~\ref{table:ex4.2} summarizes the results. 
\begin{table}[!ht]
\begin{center}
\scalebox{0.9}{
\begin{tabular}{l|c|c|c}
\hline
     $1-\alpha$ & Oracle upper bound  & Upper bound of $[0, Y_{(r)})$ & Upper bound  of $I_{\alpha}^{h}(X_{n+1}, Y^n)$   \\
 \hline
             $90\%$ & 8.08  	& 8.09    & 8.33 \\
              $92.5\%$  & 10.19   &  10.20  & 10.22\\
	   $95\%$   &15.51   	&  16.30  & 16.33 \\
	  $97.5\%$  & 33.55	&   33.63  & 33.66 \\
\hline
\end{tabular}
}
\end{center}
\caption{\small The oracle upper bound and upper bounds of the $100(1-\alpha)\%$ conformal prediction intervals $[0, Y_{(r)})$ and  $I_{\alpha}^{h}(X_{n+1}, Y^n)$, based on automobile bodily injury claims data in Example~4,  for various confidence levels. }
\label{table:ex4.2}
\end{table}
\smallskip

We see that the prediction interval $[0, Y_{(r)})$ is slightly more conservative than the oracle prediction interval. This is expected and consistent with our observations in Examples~1-3. The upper bounds of both $[0, Y_{(r)})$ and $I_{\alpha}^{h}(X_{n+1}, Y^n)$ are comparable at all chosen confidence levels, with $[0, Y_{(r)})$ being slightly shorter.  We must interpret the results here cautiously. Since we only have one sample, we cannot calculate the mean lengths of $[0, Y_{(r)})$ and $I_{\alpha}^{h}(X_{n+1}, Y^n)$. Therefore, we cannot assert that $I_{\alpha}^{h}(X_{n+1}, Y^n)$ is slightly more conservative than $[0, Y_{(r)})$.  Even if that is the case,  this slight conservativeness of $I_{\alpha}^{h}(X_{n+1}, Y^n)$ is likely due to two factors: (i) the distribution of $Y$ is heavy-tailed, as we see from Table~\ref{table:ex4.1}, and (ii)  most predictors are categorical,   adding another layer of challenge.  The excellent performance of the proposed method in Examples~1-3 makes us inclined to believe both $[0, Y_{(r)})$ and $I_{\alpha}^{h}(X_{n+1}, Y^n)$ are all finite-sample valid and efficient.

\section{Concluding remarks}

When it comes to predicting future insurance claims,  a finite-sample prediction interval is guaranteed to achieve the nominal probability level.  However,  the current literature shows a dearth of implementable finite-sample prediction intervals in the regression setting.  To alleviate this issue,  this article proposes a new strategy that converts a predictive method in the unsupervised iid setting to a predictive method in the regression setting.  In particular, this new strategy allows us to derive infinitely many finite-sample valid prediction intervals in the regression setting.  Though we have focused on interval prediction in this article, the proposed strategy applies to point prediction.  To see that, we first write (\ref{eq:true}) as (\ref{eq:true2}), i.e., 
\[
Y= h(X)+W, 
\]
where $W=Y-h(X)$ and $h$ is a  transformation.  Next, we can apply one of the many existing methods for point prediction in the unsupervised iid setting (e.g.,  Jeon and Kim 2013; Hong and Martin 2017; Lee and Lin 2010) to obtain a point prediction $\widehat{W_{n+1}}$  of $W_{n+1}$,  based on $W^n=\{W_1, \ldots, W_n\}$. Finally, we predict $Y_{n+1}$ as 
\begin{equation}
\label{eq:pointpred}
\widehat{Y_{n+1}}=h(X_{n+1})+\widehat{W_{n+1}}.
\end{equation}
Also, the assumption $E[\varepsilon]=0$ adopted for (\ref{eq:true}) is conventional but not necessary.  A scrutiny of Section~3 and the Appendix shows that our method goes through without this constraint.



\section*{Conflicts of interest}
The author has no conflicts of interest to declare.

\section*{Appendix}

\subsection*{A review of conformal prediction}

Conformal prediction is a general learning strategy for constructing finite-sample valid prediction regions; see Shafer and Vovk (2008) for an introduction and Vovk et al. (2005) for a comprehensive treatment.  Conformal prediction can be applied in both supervised learning and unsupervised learning.  We focus on supervised learning only.

To construct a conformal prediction region,  we start with a real-valued deterministic mapping $M$ with two arguments $(B, z)\mapsto M(B, z)$, where the first argument $B$, called a \emph{bag},  is a collection of observed data,  and the second argument $z$ is a provisional value of a future observation.  The mapping $M$, called a \emph{non-conformity measure}. It measures the degree of non-conformity of the provisional value $z$ with the data in the bag $B$.  If $M(B, z)$ is relatively small,  we interpret it that $z$ agrees with the data in $B$;  if $M(B, z)$ is relatively large,  we believe $z$ does not comport with the data in $B$.  There is no unique choice of the non-conformity measure.  For example,  if $z$ is real-valued and $B=\{z_1, \ldots, z_n\}$, then we can take $M(B, z)=|\hat{m}_B(x)-y|$, where $\hat m_B(\cdot)$ is an estimate of the conditional mean function, $\E(Y \mid X=x)$, based on the bag $B$.  In particular, if a linear model is taken and the least squares method is employed, then  $\hat m_B(\cdot)$ is the absolute least squares residual; if a linear model is fitted using lasso, then 
$\hat m_B(\cdot)$ is the absolute lasso residual.  In practice,  the non-conformity is chosen at the discretion of the actuary.  Once a non-conformity measure is selected, the actuary runs the conformal prediction algorithm---Algorithm~\ref{algo:conformal.s} to assess the plausibility of the value $y$ of the next label $Y_{n+1}$ at a randomly sampled feature $X_{n+1}=x_{n+1}$.

\begin{algorithm}
Initialize: data $z^n = \{z_1,\ldots,z_n\}$ and $x_{n+1}$, non-conformity measure $M$, and a possible $y$ value\;
Put $z_{n+1} = (x_{n+1}, y)$ and $z^{n+1} = z^n \cup \{z_{n+1} \}$\;
Define $\mu_i = M(z^{n+1} \setminus \{z_i\}, z_i)$ for $i=1,\ldots,n,n+1$\;
Compute $\pl_{z^n}(x_{n+1}, y) = (n+1)^{-1} \sum_{i=1}^{n+1} 1\{\mu_i \geq \mu_{n+1}\}$\;
Return $\pl_{z^n}(x_{n+1}, y)$\;
\caption{\bf Conformal prediction (supervised learning)}
\label{algo:conformal.s}
\end{algorithm}

In Algorithm~\ref{algo:conformal.s}, $1_A$ denotes the indicator function of an event $A$, i.e., 
\[
1_A(x)=\left\{ \begin{array}{ll}
		                          1,& \hbox{if  $x\in A$,} \\
		                          0,& \hbox{if $x\not\in A$.} \\
		                          \end{array}
		                         \right.
\]
The quantity $\mu_i$ is called the $i$-th \emph{non-conformity score}. It 
assigns a numerical value in $[0, 1]$ to $z_i$ to measure of agreement between $z_i$ with the data in the $i$-th augmented bag $\widetilde{B}_i=z^n\cup\{z_{n+1}\}\backslash\{z_i\}$, where $z_i$ itself is excluded to avoid biases as in leave-one-out cross-validation.  Algorithm~\ref{algo:conformal.s} corresponds to the function $\pl_{z^n}$: for a given $x_{n+1}$ and a provisional $y$, it outputs $\pl_{z^n}(x_{n+1}, y)$. The function $\pl_{z^n}$,  called  the \emph{plausibility function}, indicates how plausible $z$ is a value of $Z_{n+1}$, based on the available data $Z^n=z^n$ and $X_{n+1}=x_{n+1}$ by outputting a value between $0$ and $1$.  Using the output of the plausibility function $\pl_{z^n}$,  the actuary can construct a $100(1-\alpha)\%$ conformal prediction region
\begin{equation}
\label{eq:region}
C_\alpha(x; Z^n) = \{y: \pl_{Z^n}(x, y) > \alpha\},
\end{equation}
where $0<\alpha<1$.  
One of the key advocated advantages of conformal prediction is its use for creating finite-sample valid prediction regions. 
The following Theorem,  proved in Hong (2026),  establishes the finite-sample validity of $C_\alpha(x; Z^n)$.

\begin{thm}
\label{thm:valid}
Let $\prob$ denote the distribution of an exchangeable sequence $Z_1,Z_2,\ldots$,  and let $\prob^{n+1}$ be the corresponding joint distribution of $Z^{n+1}=\{Z_1,\ldots,Z_n,Z_{n+1}\}$.  For $0<\alpha<1$, put $t_n(\alpha) = (n+1)^{-1}\lfloor (n+1)\alpha \rfloor$, where $\lfloor a \rfloor$ denotes the greatest integer less than or equal to $a$.  Then
\begin{equation*}
\sup \prob^{n+1}\{ \pl_{Z^n}(Z_{n+1}) \leq t_n(\alpha) \} \leq \alpha \quad \text{for all $n$ and all $\alpha \in (0,1)$},
\end{equation*}
where the supremum is over all distributions $\prob$ for the exchangeable sequence.
\end{thm}

\subsection*{A subtle issue in the current literature on conformal prediction}

There is a subtle issue in the current literature on conformal prediction.   To see this, we notice that nearly all papers write the conformal prediction algorithm in the following manner:\\

\begin{algorithm}[H]
Initialize: data $z^n = \{z_1,\ldots,z_n\}$ and $x_{n+1}$, non-conformity measure $M$, and a grid of $y$ value\;
\For{each possible $y$} {
Set $z_{n+1} = (x_{n+1}, y)$ and write $z^{n+1} = z^n \cup \{z_{n+1} \}$\;
Define $\mu_i = M(z^{n+1} \setminus \{z_i\}, z_i)$ for $i=1,\ldots,n,n+1$\;
Compute $\pl_{z^n}(x_{n+1}, y) = (n+1)^{-1} \sum_{i=1}^{n+1} 1\{\mu_i \geq \mu_{n+1}\}$\;
}
Return $\pl_{z^n}(x_{n+1}, y)$ for each  $y$ on the grid\;
\caption{\bf Conformal prediction (supervised learning)}
\label{algo:conformal.s2}
\end{algorithm}

\bigskip

Indeed,  only Hong and Nasreddine (2025) and Hong (2026) depart from the prior literature and start to state the conformal prediction algorithm as Algorithm~\ref{algo:conformal.s}.  Algorithm~\ref{algo:conformal.s2} conceals a practical challenge: to determine the $100\%$ conformal prediction region $C_{\alpha}(x; Z^n)=\{y: \pl_{Z^n}(x, y) > \alpha\}$,  we have to calculate $\pl_{z^n}(x_{n+1}, y)$ for all possible $y$.  This cannot be realized if the response variable $Y$ takes infinitely many values (e.g., in the regression setting).  Had we used ``every possible $y$ value'' (instead of a grid of $y$ values) in Algorithm~\ref{algo:conformal.s2},  it would not be an algorithm, i.e., an effective procedure (e.g.,  Rogers 1967; Cutland 1980); it would be a non-effective procedure because it could not be completed in a finite time.  Thus,  any prediction region determined using Algorithm~\ref{algo:conformal.s2} is not the $100(1-\alpha)\%$ conformal prediction region $C_\alpha(x; Z^n)$ given by (\ref{eq:region}), but an approximation $\widehat{C_\alpha(x; Z^n)}$ to it.  Note that finite-sample validity of $\widehat{C_\alpha(x; Z^n)}$ is nowhere justified,  though $C_\alpha(x; Z^n)$ is always finite-sample valid.  Algorithm~\ref{algo:conformal.s} sheds light on the conformal prediction algorithm: it corresponds to the function $\pl_{z^n}(x_{n+1}, \cdot)$.  It also highlights the key challenge in applying conformal prediction: we need to calculate $\pl_{z^n}(x_{n+1}, y)$ for all possible $y$ when determining the $100\%$ conformal prediction region $C_{\alpha}(x; Z^n)$. This challenge cannot be overcome by brute force since real-world computation must be finished in finite steps.  One feasible strategy is to show that, for a particular non-conformity measure,  the conformal prediction region $C_\alpha(x; Z^n)$ has an equivalent form that can be determined exactly in finite steps. This is the strategy used in Hong and Martin (2021), Hong and Nasreddine (2025),  and Hong (2026).  Therefore, all conformal prediction regions in these three papers are finite-sample valid and ``safe'' to use.  In particular,  the $100(1-\alpha)\%$ conformal prediction $I^h_{\alpha}(X_{n+1}, Y^n)$ proposed in this article is based on the conformal prediction interval in Hong and Martin (2021). Therefore, it is genuinely finite-sample valid too.

\section*{References}
\begin{description}


\item{} Brazauskas, V.~and Kleefeld, A.~(2011). Folded and log-folded-$t$ distributions as models for insurance loss data. \emph{Scandinavian Actuarial Journal} 1, 59--74.

\item{} Brazauskas, Y.~and Kleefeld, A.~(2016). Modeling severity and measuring tail risk of Norwegian fire claims. \emph{North American Actuarial Journal} 20(1), 1--16.


\item{} Calder\'{\i}n-Ojeda, E.~and Kwok, C.~F.~(2016). Modeling claims data with composite Stoppa models. \emph{Scandinavian Actuarial Journal} 9, 817--836.

\item{} Cooray, K.~and Ananda, M.~A.~M.~(2005). Modeling actuarial data with a composite lognormal-Pareto model. \emph{Scandinavian Actuarial Journal}~5, 321--334.

\item{} Cutland, N. ~(1980). \emph{Computability: An introduction to recursive function theory}. Cambridge University Press: Cambridge, UK. 



\item Fellingham, G.W., Kottas, A. and Hartman, B.~(2015). Bayesian non-parametric predictive modeling of group health claims. \emph{Insurance: Mathematics and Economics}, 60, 1--10.

\item{} Frees, E. W.~(2010). \emph{Regression Modeling with Actuarial and Financial Applications}, Cambridge: Cambridge University Press.

\item{} Frees, E.~W., Derrig, R.~A.~and Meyers, G.~(2014). \emph{Predictive Modeling Applications in Actuarial Science, Vol. I: Predictive Modeling Techniques}, Cambridge: Cambridge University Press.

\item{} Frey, J.~(2013). Data-driven non-parametric prediction intervals. \emph{Journal of Statistical Planning and Inference}~143, 1039--1048.



\item{} Hong, L.~(2026).  Conformal prediction of future insurance claims in the regression problem. \emph{European Actuarial Journal}, to appear, \url{https://arxiv.org/abs/2503.03659}.

\item{} Hong, L., Kuffner, T. ~and Martin, R. ~(2018a). On overfitting and post-selection uncertainty assessments.  \emph{Biometrika}~105(1), 221--224.

\item{} Hong, L., Kuffner, T. ~and Martin R. ~(2018b). On prediction of future insurance claims when the model is uncertain. \emph{Variance}~12(1), 90--99.

\item{} Hong, L.~and Martin, R.~(2017). A flexible Bayesian non-parametric model for predicting future insurance claims.  \emph{North American Actuarial Journal}~21(2), 228--241.  

\item{} Hong, L. and Martin, R. (2020). Model misspecification, Bayesian versus credibility estimation, and Gibbs posteriors. \emph{Scandinavian Actuarial Journal}~2020(7), 634--649.

\item{} Hong, L.~and Martin, R.~(2021). Valid model-free prediction of future insurance claims. \emph{North American Actuarial Journal}~25(4), 473--483.

\item{} Hong, L.~and Nasreddine, N.R.~(2025). On some practical challenges of conformal prediction,  under review, \url{https://arxiv.org/abs/2510.10324}.

\item Jeon, Y.~and Kim, J.~H.~T.~(2013). A gamma kernel density estimation for insurance loss data. \emph{Insurance: Mathematics and Economics}~53, 569--579.

\item{} Kuchibhotla, A.K. Kolassa, J.E.~and Kuffner, T.A.~(2022). Post-Selection Inference. \emph{Annual Review of Statistics and Its Application}~9,  505--527.

\item{} Lee, S.~C.~K.~and Lin, X.~S.~(2010). Modeling and evaluating insurance losses via mixtures of Erlang distributions, \emph{North American Actuarial Journal}~14 (1), 107--130.

\item{} Lei, J., Robins, J.~and Wasserman, L.~(2013). Distribution-free prediction sets. \emph{Journal of American Statistical Association}~108(501), 278--287.

\item{} Lei, J.~and Wasserman, L.~(2014). Distribution-free prediction bands for non-parametric regression.  \emph{Journal of Royal Statistical Society}-Series~B~(76), 71--96.

\item{} Lei, J.,  G'Sell, M., Rinaldo, A., Tibshirani, R.J.~and Wasserman, L.~(2018). Distribution-free predictive inference for regression.  \emph{Journal of the American Statistical Association}~113(523), 1094--1111.

\item{} Nadarajah, S.~and Bakar, S.~A.~A.~(2014). New composite models for the Danish fire insurance data. \emph{Scandinavian Actuarial Journal} 2, 180--187.


\item{} Pigeon, M.~and Denuit, M.~(2011). Composite lognormal-Pareto model with random threshold. \emph{Scandinavian Actuarial Journal}~3, 177--192.

\item{} Richardson, R.~and Hartman, B.~(2018). Bayesian non-parametric regression models for modeling and predicting healthcare claims. \emph{Insurance: Mathematics and Economics}~83, 1--8.

\item{} Rogers, J.H.~(1967). \emph{Theory of Recursive Functions and Effective Computability}.  Mc Graw-Hill: New York.

\item{} Scollnik, D.~P.~M.~(2007). On composite lognormal--Pareto models. \emph{Scandinavian Actuarial Journal}~1: 20--33.


\item{} Shafer, G.~and Vovk, V.~(2008). A tutorial on conformal prediction. \emph{Journal of Machine Learning}~9, 371--421.

\item{} Shmueli, G.~(2010). To explain or predict? \emph{Statistical Sciences}~25(3), 289--310. 

\item{} Tian Q, Nordman D.J.~and Meeker W.Q.~(2022).  Methods to compute prediction intervals: a review and new results. \emph{Statistical Sciences}~37(4), 580--597. 


\item{} Tukey, J.W.~(1962). The future of data analysis. \emph{Annals of Mathematical Statistics}~33, 1--67. 

\item{} Vovk, V., Gammerman, A., and Shafer, G.~(2005). \emph{Algorithmic Learning in a Random World}. New York: Springer.



\end{description}

\end{document}